\documentclass[aps,prl,preprint,showpacs]{revtex4}
\usepackage{amsmath}
\usepackage{amsfonts}
\usepackage{amssymb}
\begin{document}
\title{On the relationship between quantum entanglement \\[-3pt] and classical
synchronization in open systems}
\author{E.D. Vol}
\affiliation{B. Verkin Institute for Low Temperature Physics and Engineering
of the National\\ Academy of Sciences of Ukraine, 47 Lenin Avenue, 61103
Kharkov, Ukraine\vspace*{2cm}}

\begin{abstract}
\vspace*{0.2cm}
We propose a simple model of classical open system consisting
of two subsystems all stationary states of which correspond to phase
synchronization between the subsystems. The model is generalized to quantum
systems in a finite-dimensional Hilbert space. The analysis of the simplest
two qubit version of the quantum model shows that all its stationary states
are nonseparable.
\end{abstract}
\maketitle

\emph{Entanglement} of quantum states is a notable resource of quantum
information science and plays a key role in it's advanced applications such
as superdense coding, quantum teleportation and quantum cryptography.
\cite{1}. The important question of whether there is any analogue of the
entanglement in classical systems remains open. This problem has been
considered by different approaches \cite{2,3,4}, but all of them focused on
\emph{closed} Hamiltonian systems only and tried to find a connection between
the entanglement of ground state of the system and peculiarities in its phase
portrait under the variation of parameters.

The goal of the present paper is to establish a relationship between the
entanglement of stationary states of an \emph{open} quantum system and such
well-known classical nonlinear phenomenon as \emph{synchronization}. As a
first step on this way, we consider a simple model of phase synchronization
of two subsystems which can be generalized to a quantum case. The analysis of
the simplest two qubit version of the model leads to the main result of the
paper: All stationary states of the system turns out to be entangled.
Although this result is obtained only for a particular model and has not been
generally proved, it suggests a new promising perspective for better
understanding of mixed entangled states properties.

Consider a simple dynamical model of open system with three variables $l_{x},
l_{y}$, and $l_{z}$ whose time evolution is governed by the following coupled
equations
\begin{equation}
\begin{array}{rl}
\displaystyle\frac{dl_x}{dt}= & 2\left( l_{y}^{2}+l_{z}^{2}\right),\\[11pt]
\displaystyle\frac{dl_y}{dt}= & -2l_xl_y,\\[11pt]
\displaystyle\frac{dl_z}{dt}= & -2l_xl_z.
\end{array} \label{1}
\end{equation}
This system may be considered as the model for phase synchronization of two
subsystems for the following reasons. First, remind that synchronization in
nonlinear dynamics is a phenomenon of rhythms adjustment of oscillating
systems due to weak interaction between them \cite{5}. Here we consider only
the simplest case of the phase synchronization when the phase difference of
two subsystems vanishes with time. Using phases of subsystems $\varphi_1$,
$\varphi_2$ and their amplitudes $r_1, r_2$ as dynamical variables, we can
write the complete system of equations in the form: $d\varphi_i/dt=f_i(
\varphi_1,\varphi_2,r_1,r_2)$ and $dr_i/dt=g_i(\varphi_1,\varphi_2,r_1,r_2)$,
where $i=1,2$ and $f_i$ and $g_i$ are nonlinear functions providing the
synchronization. Then, this system of four equations can be rewritten as a
system of two equations by introducing complex variables $z_1=r_1\exp i
\varphi_1$ and $z_2=r_2\exp i\varphi_2$. Finally, with new variables $l_x,
l_y,l_z$ connected to $z_1,z_2$ by the relations
\begin{equation}
l_x=\frac{z_{1}^{\ast}z_2+z_1z_{2}^{\ast}}{2}, \quad l_y=\frac{i(z_1z_{2}^
{\ast} -z_{1}^{\ast}z_2)}{2}, \quad l_z=\frac{\vert z_1\vert ^2-\left\vert z
_2\right\vert ^{2}}{2}, \label{2}
\end{equation}
we eventually arrive at a system of nonlinear differential equations for $l_x,
l_y$, and $l_z$. Note that the transformation defined by Eq.\ (\ref{2}) is
a classic analogue of the Schwinger transformation in quantum mechanics which
used for the description of two Bose oscillators in terms of angular momentum
components. Since $l_y$ can be expressed as $l_y=r_1 r_2\sin(\varphi_2-
\varphi_1)$ we assume that vanishing of $l_y$ is the condition of the phase
synchronization between the subsystems. Returning to system (\ref{1}), we
note that it has two integrals of motion: $l^2=l_{x}^{2}+l_{y}^{2}+l_{z}^{2}$
and $k=l_y/l_z$. A simple analysis shows that under arbitrary value of $l^2$
the solution of Eq.\ (\ref{1}) tends to the final state $l_y=l_z=0$ and $l_x=
l$. Thus the presence of phase synchronization in the classical model
represented by Eq.\ (\ref{1}) is proved.

Now let us discuss the question about possible quantum analogues of model
(\ref{1}). The proper way to quantize Eq.\ (\ref{1}) (at least in
semiclassical approximation) has been proposed by the author in Ref.\
\cite{6}. In the present case this method of quantization can be formulated
as follows. First, we have to represent the classic equations for components
$l_x,l_y,l_z$ in the form allowing quantization:
\begin{equation}
\frac{d\mathbf{l}}{dt}=-\left(\mathbf{l}\times\frac{\partial H_0}{d\mathbf{l}
}\right)+\mathbf{l}\times \left( iR\frac{\partial R^{\ast}}{d\mathbf{l}}-iR^{
\ast}\frac{\partial R}{\partial \mathbf{l}}\right),
\label{3}
\end{equation}
where $H_0$ is real and $R$ and $R^{\ast}$ are complex functions of $l_x,
l_y,l_z$ (star means complex conjugation). Then, following Ref.\ \cite{6},
a self-consistent quantum version of Eq.\ (\ref{3}) can be obtained if one
writes down the Lindblad equation for density matrix of the system:
\begin{equation}
\frac {d\hat{\rho}}{dt}=-i\left[\hat{H}_{0},\hat{\rho}\right]+\left[ \hat{R}
\hat{\rho},\hat{R}^{+}\right]+\left[ \hat{R},\hat{\rho}\hat{R}^+\right],
\label{4}
\end{equation}
where $\hat{H}_{0}$ and $\hat{R},\hat{R}^{+}$ are operator analogues of
functions $H_{0},R,R^{\ast}$ in Eq.\ (\ref{3}).

Certainly  the question regarding the order of operators $l_i$ arrangement in
$H_0$ and $R$ exists because of their noncommutativity, but in semiclassical
approximation this problem does not arise. It is easy to check by direct
verification that system (\ref{1}) can be represented in the form required
for quantization of Eq.\ (\ref{3}) by substitution $H=f\left( l^{2}\right),
R=l_z-il_y$ (the first term in r.h.s. of Eq.\ (\ref{3}) vanishes in this
case). Thus we see that possible quantum analogues of classical model of
synchronization (\ref{1}) can be described by the Lindblad equation
\begin{equation}
\frac{d\hat{\rho}}{dt}=\left[ \hat{R}\hat{\rho},\hat{R}^{+}\right]+\left[
\hat{R},\hat{\rho}\hat{R}^{+}\right], \label{4a}
\end{equation}
where $\hat{R}=\hat{l}_z-i\hat{l}_y$.

It should be noted that using the evolution equation for the density matrix,
given Eq.\ (\ref{4a}), we can write equations of motion for the average value
$\left\langle A\right\rangle$ of an arbitrary observable $A$. For example, if
one takes $l_x$ as $A$, the result is
\begin{equation}
\frac{d\left\langle\hat{l}_x\right\rangle }{dt}=\left\langle \hat{R}^{+}
\left[\hat{l}_x,\hat{R}\right]\right\rangle +c.c.=\left\langle\left(\hat{l}
_z+i\hat{l}_y\right)\left(\hat{l}_z-i\hat{l}_y\right)\right\rangle + c.c.
\label{5}
\end{equation}

We see that in semiclassical approximation Eq.\ (\ref{5}) coincides with the
first from Eqs.\ (\ref{1}) as it should be. Our next task is to find
stationary solutions of Eq.\ (\ref{4a}) and to evaluate their entanglement.
The simplest situation in which this can be done exactly is a two qubit
realization of Eq.\ (\ref{4}). In this case operators $l_x,l_y,l_z$ have the
following representation: $l_x=\frac 12(\sigma_x\otimes 1+1\otimes\sigma_x),
l_y=\frac 12 \left(\sigma_y\otimes 1+1\otimes\sigma_ y\right), l_z=\frac
12\left(\sigma_z \otimes1+1\otimes\sigma_z\right)$, where $\sigma
_x,\sigma_y,\sigma_z$ are ordinary Pauli matrices. Operator $\hat{l}^{2}=
\hat{l}_{x}^{2}+\hat{l}_{y} ^{2}+\hat{l}_{z}^{2}$ has the form $\hat{l}^{2}
=\frac 12\left( 3+\sigma_x \otimes\sigma_x+\sigma_y\otimes \sigma_y+\sigma
_z\otimes\sigma_z\right)$ and commutes with all $\hat{l}_i$. Operator
$\hat{R}=\hat{l}_{z} - i\hat{l}_{y}$ has the following matrix representation:
\begin{equation}
\hat{R}=\frac 12
\begin{pmatrix}
2 & -1 & -1 & 0\\
1 & 0 & 0 & -1\\
1 & 0 & 0 & -1\\
0 & 1 & 1 & -2
\end{pmatrix}. \label{5a}
\end{equation}
With the help of Eq.\ (\ref{5a}) we can find that equation $\hat{R}\left\vert
\Psi\right\rangle =0$ has two linearly independent solutions:
$$
\left\vert \Psi_1\right\rangle =\frac 12
\begin{pmatrix}
1\\[-5pt]
1\\[-5pt]
1\\[-5pt]
1
\end{pmatrix}
\qquad \text{and} \qquad
\left\vert \Psi_2\right\rangle =\frac 1{\sqrt{2}}
\begin{pmatrix}
0\\[-5pt]
1\\[-5pt]
-1\\[-5pt]
0
\end{pmatrix}.
$$
Therefore in the two qubit case the general stationary solution of Eq.\
(\ref{4a}) can be written as
\begin{equation}
\hat {\rho}_{st}=a\left\vert \Psi_1\right\rangle\left\langle\Psi_1\right\vert
+b\left\vert\Psi_2\right\rangle\left\langle\Psi_2\right\vert +c\left\vert\Psi
_1\right\rangle \left\langle \Psi_2\right\vert +h.c. \label{6}
\end{equation}
Coefficients $a,b$, and $c$ in Eq.\ (\ref{6}) must satisfy two conditions:
$a+b=1$ and $ab\geq c^2$ which correspond to normalization and positivity of
matrix $\hat{\rho}_{st}$. Using Eq.\ (\ref{6}) we can write the desired
expression for $\hat{\rho}_{st}$ as follows
\begin{equation}
\hat{\rho}_{st}=
\begin{pmatrix}
\displaystyle\frac a4&\phantom{x}&\displaystyle\frac a4+\frac{c}{2\sqrt{2}}&
\phantom{x} &\displaystyle\frac a4-\frac{c}{2\sqrt{2}}&\phantom{x}&
\displaystyle\frac a4\\[9pt]
\displaystyle\frac a4+\frac{c}{2\sqrt{2}}&\phantom{x}&\displaystyle\frac a4
+\frac b2+\frac{c}{\sqrt{2}} &\phantom{x} & \displaystyle\frac a4-\frac b2&
\phantom{x}&\displaystyle\frac a4+\frac{c}{2\sqrt{2}}\\[9pt]
\displaystyle\frac a4-\frac{c}{2\sqrt{2}} &\phantom{x} &\displaystyle\frac a4
-\frac b2&\phantom{x}&\displaystyle\frac a4+\frac b2-\frac{c}{\sqrt{2}}&
\phantom{x}&\displaystyle \frac a4-\frac{c}{2\sqrt{2}}\\[9pt]
\displaystyle\frac a4 &\phantom{x} &\displaystyle\frac a4+\frac c{2\sqrt{2}}&
\phantom{x}&\displaystyle\frac a4-\frac{c}{2\sqrt{2}}&\phantom{x}&
\displaystyle\frac a4
\end{pmatrix}. \label{6a}
\end{equation}

It is easy to check that $\hat{\rho}_{st}$ has two null eigenvalues and two
positive ones, $\lambda_1$ and $\lambda_2$, which satisfy to the following
equation: $\lambda^2-\lambda+ab-c^2=0$. Let us find the values of coefficients
$a,b,c$, for which density matrix $\hat{\rho}_{st}(a,b,c)$ corresponds to
entangled states of two qubits. A simple way to accomplish this is to invoke
the Peres criterium \cite{7}. As well known \cite{8}, in the two qubit case
this criterium is a necessary and sufficient condition of the mixed states
separability. According to it density matrix $\hat{\rho}$ is separable if
matrix $\hat{\rho}_{PT}$ obtained from $\hat{\rho}$ by operation of partial
transposition (which corresponds to permutations of indices of one of the
subsystems only) is nonnegative. Using Eq.\ (\ref{6a}) for $\hat{\rho}_{st}$
we can write $\left( \hat{\rho}_{st}\right)_{PT}$ as
\begin{equation}
\left(
\hat{\rho}_{st}\right)_{PT}=
\begin{pmatrix}
\displaystyle\frac a4 &\phantom{xx}\displaystyle\frac a4+\frac{c}{2\sqrt{2}}&
\phantom{xx}\displaystyle\frac a4-\frac c{2\sqrt{2}}&\phantom{xx}
\displaystyle\frac a4-\frac b2\\[8pt]
\displaystyle\frac a4+\frac{c}{2\sqrt{2}}&\phantom{xx} \displaystyle\frac a4
+\frac b2+\frac{c}{\sqrt{2}}&\phantom{xx}\displaystyle\frac a4 &\phantom{xx}
\displaystyle\frac a4\frac{c}{2\sqrt{2}}\\[8pt]
\displaystyle\frac a4-\frac{c}{2\sqrt{2}} &\phantom{xx}\displaystyle\frac a4&
\phantom{xx}\displaystyle\frac a4+\frac b2-\frac{c}{\sqrt{2}}&\phantom{xx}
\displaystyle\frac a4-\frac{c}{2\sqrt{2}}\\[8pt]
\displaystyle\frac a4-\frac b2 & \phantom{xx}\displaystyle\frac a4+\frac{c}{2
\sqrt{2}} &\phantom{xx}\displaystyle\frac a4-\frac{c}{2\sqrt{2}}&
\phantom{xx}\displaystyle\frac a4
\end{pmatrix}. \label{6b}
\end{equation}

It is easy to check that $\left( \hat{\rho}_{st}\right)_{PT}$ has eigenvector
$$
\left\vert \lambda_1\right\rangle =\frac {1}{\sqrt{2}}
\begin{pmatrix}
1\\[-4pt]
0\\[-4pt]
0\\[-4pt]
-1
\end{pmatrix},
$$
with eigenvalue $\lambda_1=b/2$. The remaining eigenvalues are found from
characteristic equation
\begin{equation}
\left|
\begin{array}[c]{lllll}
\displaystyle\frac{a-b}{2}-\lambda &\phantom{x} &\displaystyle\frac a4+
\displaystyle\frac{c}{2\sqrt{2}} &\phantom{x}&\displaystyle\frac a4-\frac
{c}{2\sqrt{2}} \\[6pt]
\displaystyle\frac a2+\frac{c}{\sqrt{2}} &\phantom{x} &\displaystyle\frac a4+
\frac{c}{\sqrt{2}}+\frac b2-\lambda &\phantom{x}&\displaystyle\frac a4\\[6pt]
\displaystyle\frac a4-\frac{c}{\sqrt{2}} &\phantom{x}&\displaystyle\frac a4&
\phantom{x} &\displaystyle\frac a4-\frac{c}{\sqrt{2}}+\frac b2-\lambda
\end{array}
\right|=0. \label{7}
\end{equation}
Computing this determinant explicitly we get the following cubic equation for
$\lambda_1, \lambda_2$, and $\lambda_3$
\begin{equation}
\lambda^3-\left( a+ \frac b2\right) \lambda^2-\left( \frac{b^{2}}{4}+ c^2-
\frac{ab}{2}\right) \lambda+\frac{b^3}{8}=0. \label{8}
\end{equation}
We see that two roots of Eq.\ (\ref{8}) are positive but the third is
negative (when $b\succ 0$). Thus for all $b\succ 0$ density matrix $\hat{\rho}
_{st}$ corresponds to nonseparable (entangled) state. This result obtained for
the simple model of synchronization together with some qualitative reasons
suggests that a relationship between the phase synchronization in classical
open system and the entanglement in its quantum analogue exists in more
general situations as well.

In conclusion, we want to examine our original model (\ref{1}) from somewhat
different point of view. Let us consider two functions of state $H=(l_{x}^{2}
+l_{y}^ {2}+l_{z}^{2})/2$ and $S=2l_x$. As follows from Eq.\ (\ref{1}) the
evolution of $H$ and $S$ satisfies two general conditions:
\begin{equation}
\frac{dH}{dt}=0 \qquad \text{and} \qquad\frac{dS}{dt}\geq 0. \label{9}
\end{equation}
It should be noted that system (\ref{1}) can be represented as:
\begin{equation}
\frac {dl_i}{dt}=\varepsilon_{ikl}\frac{\partial H}{\partial l_k} A_l,
\label{10}
\end{equation}
where $A_l=\varepsilon_{lmn}\partial S/\partial l_m \partial H/\partial l_n$
and $\varepsilon_{lmn}$ is the antisymmetric tensor of Levi-Civita.

As shown by the author \cite{9}, two conditions (\ref{9}) (in the general
case of $n$ variables $x_1,\dots, x_n$ describing the evolution of the
system) define a class of nonhamiltonian systems called quasithermodynamic
with interesting dynamical and statistical properties \cite{9}. But the
possibility of quantum quasithermodynamic systems existence was not discussed
in Ref.\ \cite{9}. The analysis presented above shows, in particular, that a
quantum analogue of system (\ref{1}) exists and it is a quasithermodynamical
system as well. One can see it directly from the fact that average values of
$\hat{H}= \hat{l}^{2}/2$ and $\hat{S}=2\hat{l}_x$ obviously satisfy to
conditions (\ref{9}). Sure, more detail analysis of quasithermodynamic
quantum systems is required, which is the subject of future work.

\acknowledgments I thank L.A. Pastur for valuable discussions of
results of this paper.

\end{document}